\documentclass[11pt]{article}
\usepackage{times}
\usepackage{latexsym}
\usepackage{graphics}
\usepackage{calc}
\usepackage{ifthen}

\newtheorem{theorem}{Theorem}
\newtheorem{proposition}[theorem]{Proposition}
\newtheorem{corollary}[theorem]{Corollary}

\begin{document}

\begin{center}

{\Large\textbf{Low Power Mesh Algorithms for Image Problems}}
\bigskip\bigskip

{\large Quentin F. Stout}
\bigskip

Computer Science and Engineering\\
University of Michigan
\end{center}

\vspace{0.3in}

\centerline{\textbf{Abstract}}
\medskip

\noindent
We analyze a physically motivated fine-grained mesh-connected computer model, assuming that a word of information takes a fixed area and that it takes unit time and unit energy to move a word unit distance.
This is a representation of computing on a chip with myriad tiny processors arranged as a mesh.
While most mesh algorithms assume all processors are active at all times, we give algorithms that have only a few processors on at any one time, which reduces the power required.
We apply this approach to basic problems involving images, showing that there can be dramatic reductions in the peak power with only small, if any, changes in the time required.
We also show that these algorithms give a more efficient way to utilize power when more power is available.
\bigskip


\section{Introduction}

Parallel systems connected in a grid fashion have occurred throughout computing
history, going back before 1950 with the cellular automata model that arose
through the interaction of von Neumann and Ulam~\cite{vonNeumannCA,Ulam74}.
Models of processors connected in a grid include cellular
automata~\cite{Beyer69,Burks70,Jakubowskietal01,Levialdi72,vonNeumannCA,Ulam74,Umeo01},
fine-grained SIMD parallel computers~\cite{BatcherMPP,DuffWatson77,Unger58},
VLSI designs~\cite{Ullman84}, mesh-connected 
computers~\cite{Leighton92,MillerStout89,MIT,NassimiSahni80,ThKu77},  
sensor networks (typically an irregular grid)~\cite{Tim2018,SiBaPrasanna04}, 
multicore chips~\cite{Tilera},
and supercomputers~\cite{BlueGene,IntelTFLOPS}.
For such systems near-neighbor proximity
plays an important role and needs to be taken into account if one is to optimize
algorithms.  This is in contrast to typical theoretical parallel algorithms for PRAMs
which do not try to model physical aspects such as location and distance.
More recently another physical consideration, power
consumption, has taken on importance.  It is a concern in systems such as sensor networks,
cell phones, and supercomputers.
The last is different from the others in that the power is
external and does not diminish over time, but supplying the peak power
required is a major concern.  Removing the heat generated
introduces packing constraints which in turn affect communication time,
and hence various tradeoffs need to be made.
For example, peak power considerations resulted in the BlueGene systems utilizing
slow, but numerous, processors~\cite{BlueGene}, and is an important constraint in using 
nearly a million cores on a wafer-scale system~\cite{Cerebras}.

This paper addresses the problem of minimizing the peak power required by
algorithms for systems consisting of myriads of tiny, simple processors on a chip
operating in SIMD fashion.
For example, there
are image processing chips which both detect images and do substantial
processing on them. It is not realistic to assume that the entire chip can have
most of its logic circuits active at the same time, i.e., one cannot ``light up''
the entire chip, especially if it is to be used for an extended amount of time
 (e.g., see~\cite{Brown05} for an
image processing chip with processing capabilities and emphasis on energy
efficiency).
As the number of transistors per chip is increasing, the ability to have
all circuits active is decreasing.

In Section~\ref{sec:fundamentals} the basic computational model is described.
``Squirrels'' are introduced as a simple way of keeping track of processor activation and 
message passing.
Section~\ref{sec:image} contains results for images, showing that
the peak power can be reduced significantly without increasing the time required.
Section~\ref{sec:singleprocessor} examines a variant of interest in sensor networks,
where there is a limit on the total time that any processor can be active.
Section~\ref{sec:morepower} shows how the previous algorithms can be used to efficiency utilize power when more is available.
This is accomplished by using power to increase the speed of the calculations of the active processors, while still keeping most of them inactive.
Section~\ref{sec:final} contains some final remarks.

\section{Fundamentals}~\label{sec:fundamentals}

We utilize the basic fine-grained \textit{mesh-connected computer} model:
the system has
$n^2$ processors arranged as an
$n \times n$ grid, where each processor can communicate only with its
immediate neighbors (either the 4 neighbors sharing an edge, or the 8 sharing
an edge or corner).
To simplify exposition we assume that $n$ is a power of 2, with modifications
to the more general case being straightforward.
 Each processor can store a fixed number of words of
logarithmic length, and all operations on these words, including sending one to
a neighbor, take constant time and energy. Each processor starts with its
coordinates
$(x,y)$, $x, y \in [0,1,\ldots n\!-\!1]$.  
Often the processor's
\textit{z-order index} will be used, where  the processor at
$(x_kx_{k-1}\ldots x_0,\,y_ky_{k-1}\ldots y_0)$ has z-order index
$y_kx_ky_{k-1}x_{k-1}\ldots y_0x_0$.
Hilbert curve ordering would be equally useful.
For purposes of timing analysis the system is
SIMD, though strict synchronization can be relaxed.

Note that such systems are easily constructed with thousands of processors per
chip, and nearly a million in wafer-scale systems~\cite{Cerebras}.
The mesh model has long been studied and many algorithms have been developed for
it.  See, for example,~\cite{Leighton92,MIT} and the numerous references therein.

We say that a processor is \textit{active} if it is calculating or
communicating, and otherwise it is \textit{inactive}.
Inactive processors are in a very low-power state, and peak power analyses
are in terms of the number of processors in an active state.
By the parallel \textit{work} or \textit{total energy} we mean the sum, over all processors, of the time
they are active.
An active processor can determine the activity level of each of its neighbors,
and can activate an inactive neighbor.
It will be seen that all of the algorithms have the
property that a processor is only calculating in a fixed period around a
message transmission, and hence to determine the number of active processors
it will suffice to merely count processor to neighbor communication, though in some cases the message may merely be to inform the adjacent processor that it should become active.

Standard mesh-connected computer algorithms assume that all processors are active at all
times, and hence for a mesh of $n^2$ processors peak power is $\Theta(n^2)$ and
total energy is the product of time and $n^2$. 
Many basic algorithms take $\Omega(n)$ time, and thus use $\Omega(n^3)$
energy~\cite{Beyer69,Leighton92,Levialdi72,MIT,NassimiSahni80,StoutMST84,ThKu77}. 
Note that $\Omega(n)$ is a lower bound on the time for
any nontrivial problem since it is the diameter of the communication network.
As for work,  $\Omega(n^3)$ is a lower bound for operations
where $\Theta(n^2)$ values may need to be transported distance $\Theta(n)$.
Sorting is the most important such example and can be completed in
$\Theta(n)$ time~\cite{ThKu77}, and thus
reducing peak power necessarily increases time. 
Finding a minimal spanning tree when the input is unordered edges has similar behavior~\cite{StoutMST84}.
For many other problems, however, the serial work is $o(n^3)$, and when there is no
requirement that $\Theta(n^2)$ values must be transmitted distance $\Theta(n)$
there is the possibility of finishing in 
$\Theta(n)$ time and peak power $o(n^2)$.

\subsection{Squirrels}

The algorithms consist of having a subset of processors being active at some time,
where each does some calculations, activates a neighbor, passes a message to the
neighbor, and then becomes inactive.
Rather than directly indicating which processors are active at a given time, it
is useful to think of trained \textit{squirrels} traversing the mesh, where the
presence of a squirrel indicates that the processor is active and the squirrel may be carrying information.
The squirrels have a memory
of a finite number of words, they can keep track of their location, and they
can leave a finite number of words at any location.
Admittedly squirrels are an unusual computing model, but since many of the
algorithms require taking information from one place to another, and potentially leaving there
and then being able to go back and get it if needed, squirrels seem
to have the requisite skills, though their ability to be trained to cooperate
with one another remains an open question.

A squirrel carrying information
from one location to another corresponds to a sequence of calculate $\rightarrow$ 
activate neighbor $\rightarrow$ communicate $\rightarrow$ deactivate steps,
where both the number of steps (time) and total work are proportional to the
distance traveled.
Squirrel algorithms have
some similarities with pebble algorithms for automata~\cite{BlumHew67,BlumSak77,Rabin67}.
Pebbles deposited at a location are used to keep track of positional information,
and can be used to help traverse mazes and
more general graphs.  Here the problem descriptions, such as labeling
components, require the ability to store words of logarithmic size at processors, rather than
the fixed size inherent in pebble algorithms, and hence the cellular automata
model of pebble algorithms is not quite suitable.

Peak power is merely the number of squirrels, denoted $s$.
Each squirrel has a unique integer id $\in \{0,\ldots,s\!-\!1\}$ and each squirrel knows $s$ and $n$. 
For any
nontrivial problem for which every position must be visited at some point the
total energy must be $\Omega(n^2)$, and hence if time $= \Theta(n^2/s)$ then the
peak power vs.\ time tradeoff is optimal. Further, if this holds for
$s = \Theta(n)$ and information must be passed from some processor to another
one distance $\Theta(n)$ away then the time is the minimum possible no matter
how large the peak power is.  
However, Section~\ref{sec:morepower} makes an important change in this assumption, using power to increase the rate of computation.

While thinking in terms of squirrels moving around
simplifies the descriptions of many algorithms, in some cases adjacent squirrels may
stay where they are and merely exchange information, and occasionally
algorithms can result in two squirrels needing to occupy the same location at the
same time. Note that in a valid algorithm for a mesh-connected computer,
the number of squirrels per location must be a constant and cannot grow unboundedly with $n$,
but that is not a problem for any of the algorithms herein.
A squirrel can determine which adjacent locations are
occupied by squirrels, since that corresponds to an active processor determining
the activity level of its neighbors.

\subsection{Stepwise Simulation}

It is trivially true that any algorithm taking $t$ time on an
$m \times m$ mesh can be stepwise
simulated in $\Theta(tm^2/s)$ time and total energy $\Theta(tm^2)$ by
$1 \leq s \leq m^2$ squirrels, where each squirrel is responsible for stepwise
simulation of a $(m/\sqrt{s}) \times (m/\sqrt{s})$
submesh.  This fact will be used in some of the algorithms in
which a small subproblem is solved by simulating a standard (power-oblivious) mesh
algorithm.

The algorithms are described for $s = n$, and it is easy to see how to
simulate them with fewer squirrels with only linear slow-down.  In general such
simulation requires some care since in addition to the operations being
performed the simulating
squirrel must move from the location of one squirrel being simulated to the
location of another, i.e., extra time is added. Contrast this with
the result mentioned in the previous paragraph, which is analogous to
starting with an algorithm of $m^2$ squirrels which never move from their initial position,
and then simulating them by $s$ squirrels which can just do simple scans of their
subsquare to move from simulating one of the original squirrels to the next.
For the algorithms in this paper, however, the
simulation tends to be simple and details will be omitted.

A single squirrel cannot in general use an algorithm for a standard serial computer
without increasing the time required since it is constrained by the physical
dispersion of information, something normally ignored in serial algorithms
(unless cache behavior or paging is a concern).
Squirrels also introduce a constraint in that the pattern of activation in the
parallel computer follows connected paths and does not jump around.  While we
don't make this assumption for the underlying power-constrained parallel computer
model,
we don't know of problems for which this extra capability provides faster
(in O-notation) algorithms.

\section{Image Algorithms} \label{sec:image}

For image data we assume that a $n \times n$\, image is stored one
pixel per processor, where each pixel has a color.  By a \textit{figure} we
mean a connected component of pixels of the same color, where we typically
think of figures representing objects on a white background.  We
consider two pixels to be
adjacent if and only if they share an edge, but this can trivially be expanded
to include corner adjacency. By \textit{labeling figures} we mean that each
pixel is assigned a label, and that two pixels have the same label if and only
if they are in the same figure. Initially each pixel starts with its label
being its processor's z-index.  The final label of the figure will be the
minimal initial label of any of its pixels, and this position will be called
the figure's \textit{leader}.

A single squirrel can label the figures in $\Theta(n^2)$ time. To do so, it uses
a simple z-ordered scan to traverse the image.  When it encounters a pixel with
label equal to its initial label then it has encountered a new figure. It can
then use a depth-first search to label the figure in time proportional to its
size. Once the figure is labeled the squirrel returns to figure's leader and
resumes the scan.  The scan
takes
$\Theta(n^2)$ time, and since the time to label all pixels in a figure is proportional to
their number, the total labeling time is $\Theta(n^2)$, giving
$\Theta(n^2)$ time for the algorithm.
For multiple squirrels, however, a significantly different approach will be used.

\begin{figure}[t]
\centerline{\parbox{3in}{\centerline{\includegraphics{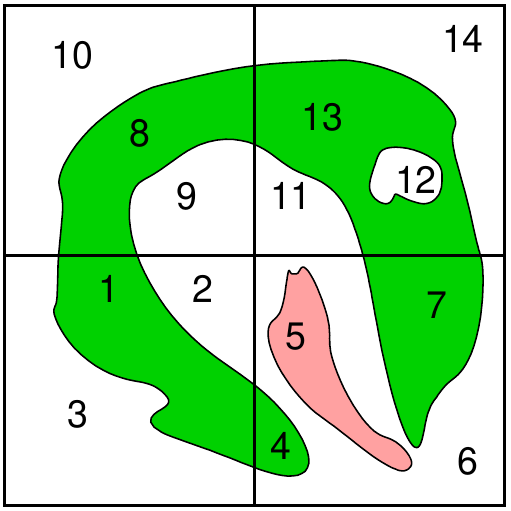}} \centerline{image}}~
                  \parbox{3in}{\centerline{\includegraphics{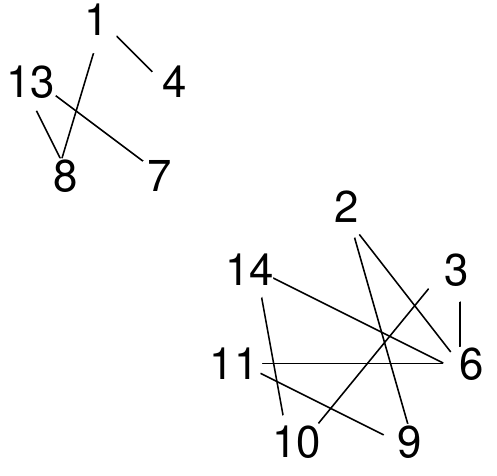}} \centerline{label graph}}}
\medskip

\caption{Merging subsquares to label figures}

\hrulefill

\label{fig:imagelabel}
\end{figure}

\begin{theorem} \label{thm:imagelabel}
Labeling the figures of an $n \times n$ image can be performed in
$\Theta((n \log s)/s)$ time using peak power $s$, for
        $1 \leq s \leq n \log n$.
\end{theorem}
\textit{Proof:\,}
A single figure may have $\Theta(n^2)$ pixels, and hence having only one squirrel work
on it would not improve upon the time of a single squirrel labeling the entire
image. Instead, we use a well-known divide-and-conquer approach (see [12,~p.30]
for a generic version of this approach).
The algorithm is as follows: to solve the problem in a square, suppose the
problem has been solved for the 4 subsquares. Within the larger square, the
only figures where the pixels' labels are inconsistent are those that cross the
borders between subsquares. The edges connecting the two sides, the ones that
contain the information needed to make the labels consistent, form the edges of
a graph in which vertices are the labels of pieces adjacent to the sides.
In this graph the connected components need to be labeled (see
Figure~\ref{fig:imagelabel}). To label this graph, move all of the edge
information to the center and use an edge-based graph algorithm to label the
components, and then move the information back to the edges. Ultimately the
highest level is reached, and then the final labels are propagated by reversing
the process.

When squares of size $m^2$ are being worked on the movement of the
subsquares' edge information to a submesh of size $\Theta(m)$ involves
$\Theta(m)$ edges being moved a distance of
$\Theta(m)$, for a total work of $\Theta(m^2)$ per square. The edge-based
component labeling involves a submesh of size $\Theta(m)$, and with a fully powered mesh can be done in
$\Theta(m^{1/2})$ time~\cite{StoutMST84} and $\Theta(m^{3/2})$ total energy,
and thus the movement to the center dominates the time and energy. Since
the movement to the center takes work proportional to the area, and the
squirrels are evenly distributed among the squares, each level of recursion
takes the same time, $\Theta(n^2/s)$.  There are $\log_2(n)-\log_2(n^2/s) =
\log_4(s)$ levels of parallel recursion, so the total time
is $\Theta((n^2 \log s)/s)$.

To finish the theorem we need to consider the power range
$n \leq s \leq n \log n$.  Partition
the mesh into subsquares of $r=(n/s)^2$ pixels.  There are $n/r$ such squares,
so each can be assigned $sr/n = n/s$ squirrels.  Note that this is $\sqrt{r}$,
and hence by the above the subsquares can be labeled in $\Theta(\sqrt{r} \log
r)$ time. Now a single merge step is used, merging all $n/r$ squares
at once. The number of squirrels is linear in the size of all of the boundaries
of the squares, so the label information of all of the boundaries can be
simultaneously moved to the center and made consistent, in
$\Theta(n)$ time.
As long as $\sqrt{r} \log r = \Omega(n)$, the time for
labeling the squares dominates the total time.  Since
$\sqrt{r} \log r = 2(n/s) \log(n/s)$, the time is as claimed for $s \leq n \log n$.
$\Box$

This divide-and-conquer approach has also been utilized for sensor network
algorithms~\cite{SiBaPrasanna04}.  In Section~\ref{sec:singleprocessor} it
will be shown that the above algorithm can be adjusted to achieve the
energy goals in~\cite{SiBaPrasanna04,Tim2018}, namely limiting the energy expended by any single processor,
while retaining its peak power properties.

Since rodents are being used to perform the algorithms, the following seems appropriate:
\begin{corollary}   \label{cor:maze}
Given an $n \times n$ black/white maze with start and stop sites,
in $\Theta((n^2 \log s)/s)$ time $s$ squirrels,
$1 \leq s \leq n \log n$,
can decide if the maze has a path from start to stop.
$\Box$
\end{corollary}
There is such a path if and only if the start and stop are in the same component.
Note that this does not say that a shortest path has been determined, merely that
they can determine if there is a path.
The power/time
tradeoff for finding a shortest path is an open question, even for a single squirrel.

\subsection{Strongly Labeled Figures}

The algorithm in Theorem~\ref{thm:imagelabel} is within a logarithmic factor of
work-optimal parallelization, and it is an open question whether this factor
can be eliminated.  Further, when $s = n$ the time is slower than the
optimal time by a logarithmic factor~\cite{NassimiSahni80}.  Once the figures
have been labeled various properties of them can be determined without the
extra logarithmic factor, but a bit more care is needed. The \textit{large}
figures, those having more than $n$ pixels, are partitioned into pieces and
the results on the pieces are combined to get the final result. The pieces are
of size $n$, and a squirrel will work on a piece and take the result directly
to a location where the piecewise results are combined all at the same time,
rather than combining them in the tree-like fashion used for labeling.  A
figure with $n$ or fewer pixels will be called \textit{small}.

\begin{figure}[t]

\centerline{\includegraphics{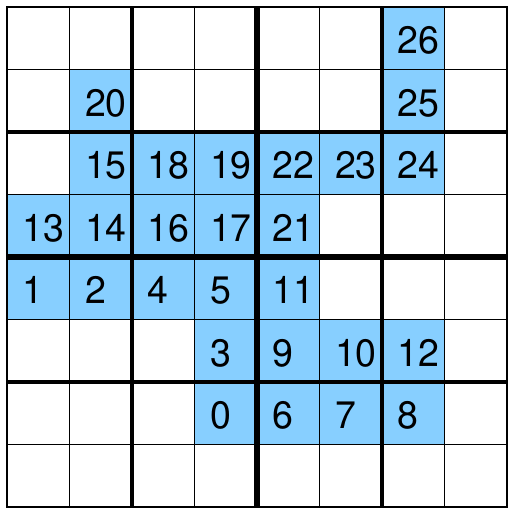}}

\caption{Rank ordering within a figure}

\hrulefill

\label{fig:rank}
\end{figure}

Within a figure, the \textit{rank} of a pixel is its position in the z-order
numbering of the pixels in the figure (with the numbering
starting at 0, i.e., the label's leader is rank 0).  See Figure~\ref{fig:rank}.
A processor in a large figure with rank a
multiple of $n$ is a \textit{breakpoint}.  We say that an image's figures are 
\textit{strongly labeled}\, if in addition to the figures being labeled, 
every processor contains the processor's rank in its figure and the
number of pixels in the figure.  Each breakpoint also contains the location
of the next breakpoint.  It
is straightforward to determine ranks as the figures are being labeled, without
increasing the time, using the property that when squares are being merged they
contain consecutive positions in the space-filling curve ordering, and hence
merely knowing the number of processors in the subsquares allows one to
determine the starting rank of processors in each subsquare.

For the processors that are breakpoints, once the ranks are known a 
simple bottom-up then top-down pass can be used to determine the location of
the next breakpoint.  A somewhat more complicated approach can
reduce the time to $\Theta($n$)$.
It introduces a technique that will be employed
in more complex operations.  Note that all of the large figures combined
have at most $2n$ breakpoints since each corresponds to a collection of $n$
pixels, except for the last breakpoint in a figure which may be in a piece containing
only itself.
Thus if $s = n$ we can assign each squirrel at most 2 breakpoints to be responsible for.
However, a given region may have many breakpoints.
To assign breakpoints to squirrels, first have the squirrels, one per row, move from right
to left, counting the number of breakpoints encountered.
Once these totals have been deposited in the leftmost column, it's simple to have each squirrel proceed
bottom-up to determine which row(s) contains its breakpoints, and then within the
row(s) determine which breakpoints it is assigned to.
Temporarily, for the purposes of locating
the next breakpoint, the pixel of highest rank within its figure is also
treated as a breakpoint, and hence there may be nearly $3 n$ points
involved, so each squirrel really carries 3 points.

Once each squirrel has located a breakpoint, it creates a record containing the
breakpoint's label, rank, and location, and then carries this to a central
$n^{1/4}\times n^{1/4}$ subsquare.  A simple mesh
algorithm is used to
sort the records by label, and within each label by rank.  If the sort is into
alternating row major order (or any other contiguous ordering) then for each
breakpoint the record of the next breakpoint in its figure is in an
adjacent record.  This information is added to the breakpoint's record, and
then the squirrels carry the records back to the breakpoints and deposit the location
of the next breakpoint, completing the operation.

By \textit{broadcast over figures} we mean that there is a value at the leader
which is then copied to every pixel in the figure. By \textit{reduction over figures}
 we mean that there is a commutative semigroup operation $\ast$ over a set
$S$, and that each pixel $p$ has a value $v(p) \in S$.  At the end of the
reduction operation, the leader of figure $F$ has the value $\ast \{v(p): p \in
F\}$.
We assume that $\ast$ can be computed in unit time. Broadcast and reduction
can easily be performed using the
divide-and-conquer approach in Theorem~\ref{thm:imagelabel}, taking the same
time bounds.  Here, however, we remove the extra logarithmic factor.

\begin{theorem} \label{thm:imagereduce}
Given a strongly component labeled $n \times n$  image, broadcast and reduction over figures
can be performed in $\Theta(n^2/s)$ time using peak power
$s$, for $1 \leq s \leq n$.
\end{theorem}
\textit{Proof:\,} 
Note that by using depth-first search a single squirrel can do broadcast and
reduction over a figure in time proportional to the size of the figure. This
will be used for small figures, and for pieces of large figures.

To do the operation for all of the small figures, first the squirrels move
right to left, one per row, with each counting the total size of all small
figures with a leader in the row.  Each total is divided by $n$ and rounded
up, with the result deposited in the leftmost column.  This the minimum number
of squirrels required if each visits no more than $n$ pixels while labeling
the small figures.  For the entire image this could require nearly $2n$
squirrels. We therefore require each squirrel to do the work of 2.  If, for
example, the first row had a total of 4.5$n$, then it was converted into a 5,
so the first 2 squirrels will work solely on that row, and the third will work
on that row and the next row with a nonzero value. Within a row the figure
sizes may not divide evenly by $n$, so the first squirrel does the first set
of figures that add up to at least 2$n$, the second squirrel takes the next
figure through the set of figures that add up to at least 4$n$, and the third
squirrel takes the remaining ones (it is possible that there are none
remaining). No squirrel works on more than 1 small figure more than its share,
and since no small figure has size more than $n$, no squirrel works on more
than 3$n$ pixels.  Hence the total time to complete the operation on all
small figures is $\Theta(n)$.

\begin{figure}[t]

\centerline{\includegraphics{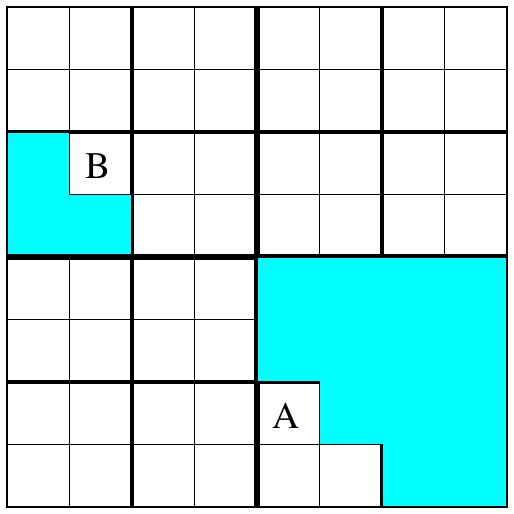}}

\caption{All locations with z-ordering between those of A and B}

\hrulefill

\label{fig:zregion}
\end{figure}

For the large figures the reduction operation will be described, with the
broadcast being an approximate reversal of this. As before we assign squirrels to
breakpoints, and the squirrel assigned to the breakpoint of rank
$in$ will do the reduction over all pixels of rank $in \ldots (i\!+\!1)n -\!1$, i.e., until the next breakpoint. 
A slight difficulty, however,
is that these pixels may not be contiguous. For example, even if the entire image is a
single figure, two pixels with consecutive ranks can
be quite far from each other because of the jumps in the z-ordering (see Figure~\ref{fig:rank}).
However, a slight modification can remedy this.  Given two pixels at positions
$p$ and $q$, the set of pixels with z-orderings between theirs form at most
2 convex regions $C_1$, $C_2$ that can easily be determined (see Figure~\ref{fig:zregion}).
If $p$ is a breakpoint of some figure $F$ and $q$ is the next breakpoint of $F$,
then all of the points in $F$ with ranks between the ranks of $p$ and $q$ lie
in $C = C_1 \cup C_2$. Further,
$F \cap C$ forms a collection of subfigures, each of which touches the
boundary of $C$.  Therefore a squirrel can start at $p$, follow the boundary of $C$,
and whenever a new pixel of $F$ is encountered start a depth-first search of
that subfigure.  The boundary of $C$ has length $O(n)$, and the total
number of pixels examined in the depth-first search is $\leq n$, so the total
time for the squirrel to determine the reduction of its piece is
$\Theta(n)$.  Then the squirrels can congregate in the middle to determine
the reductions over entire figures and return the results to the leaders. Note
that the number of pixels in $C$ may be far larger than $n$, so
the squirrel could not simply traverse all of $C$.
$\Box$

There is a slight complication in the above, in that several squirrels may have
paths that overlap, and only a fixed number are allowed to occupy a position
at any one time.  However, simple routing control mechanisms can guarantee
that all of the traversals can be completed in $O(n)$ time.

Using reduction one can find the area and perimeter of each figure, and its
bounding box, where the \textit{(iso-oriented) bounding box} of $F$, denoted
$\mathrm{box}(F)$, has x-extent equal to
the x-extent of the pixels in $F$ (i.e., the smallest to largest x-coordinates
of pixels in $F$), and its y-extent is the y-extent of the pixels in $F$.
A broadcast is used to move all of these values to all pixels in $F$.

A figure $F$ is \textit{contained in} figure
$G$ if every path from $F$ to the edge of the image contains a pixel of
$G$.  Whenever two pixels of different colors are adjacent one corresponds to a
figure containing the other (unless both figures are adjacent to the boundary),
 but it cannot be determined which is which without
some global information. Fortunately this is simple to determine since a figure
$F$ is contained in an adjacent figure $G$ iff
$\mathrm{box}(F)$ is contained in $\mathrm{box}(G)$.
Note that if figures are not adjacent
then it can be that $\mathrm{box}(F) \subset \mathrm{box}(G)$ without $G$
containing $F$.  For example, if $G$ is shaped like a U then $F$ can be a dot
inside it without being properly contained.  However, for adjacent figures this
cannot occur.
Note that a figure has only one containing figure that is adjacent to it.
The \textit{nesting level} of a figure is the number of figures that contain it.

\begin{proposition}
Given a strongly component labeled $n \times n$ image, for each figure one can determine
if it is contained in any others, obtain the label of the smallest container,
and determine its nesting level, in
$\Theta(n^2/s)$ time using peak power $s$, for $1 \leq s \leq n$.
Further, for black/white images, in the same time/power bounds each figure can
determine the smallest container of the same color.
\end{proposition}
\textit{Proof:\,} 
For each pixel $p$ on the boundary of a figure determine if an adjacent
pixel in the figure of the opposite color is part of a figure with a bounding
box containing $p$'s, and if so retain that figure's label, while otherwise
just retain an empty label.  This label is that of the smallest container of a
different color.  To find the smallest container of the same color in
a black/white image, now each boundary pixel of $F$ adjacent to
the containing figure
$G$ of the opposite color acquires the label of $G$'s container, which is the
smallest containing figure of $F$ of the same color. Note that for arbitrarily
many colors the closest enclosing figure of the same color may be many
levels away.

To determine nesting level, a left-right scan can be used, adding one every time a
transition is made from a figure to one it contains, and subtracting one
when the opposite occurs.
$\Box$

 \subsection{Closest Point Problems}

Given an image, suppose each pixel has a (possibly empty) label, not
necessarily a label of a figure. The \textit{closest similar point problem} is
to find, for every pixel with nonempty label a closest one of the same label;
the \textit{closest black point problem} is to find, for every pixel, a closest
black one; and the \textit{closest differing point problem} is to find a
closest one with a different non-empty label.

\begin{theorem} \label{thm:closestblack1infinity}
Using peak power $s$, $1 \leq s \leq n$, in $\Theta(n^2/s)$ time the closest
black point problem, and the closest differing point problem, can be solved for
the $\ell_1$ and $\ell_\infty$ metrics.
\end{theorem}
\textit{Proof:\,}
For the closest black point problem using the $\ell_1$ metric, squirrels
perform a right-left sweep in each row, leaving, at each position, the location
of the most recently encountered (hence closest) black pixel. Then a similar
left-right sweep is performed, where the closer of the black pixels in either
direction is left at each pixel. Note that for any point $p$, either the
closest black point is in the same column, or is one of the points recorded in
its column (including the points recorded at $p$ itself).  Now vertical sweeps
are done in each direction.   Suppose an upward sweep is being done.   The squirrel
remembers the location of the closest black pixel known so far.  At each
position $q$, it compares the distance to the pixel it is carrying versus the
distance to the black pixel's location stored at $q$ in the horizontal sweeps,
and it keeps the location of the closer of these two, proceeding upwards.  At
each step, when it arrives at a pixel it is carrying the location of the closest black pixel
with vertical coordinate no larger than the coordinate it is currently at.  A
similar downward sweep is also done, at which time the correct value is stored
at each location.  Modifications for the $\ell_\infty$ metric are quite simple,
and to modify for the closest differing pixel problem, note that the squirrel merely
needs to keep track of the closest labeled point, and the closest one of a
different label.
$\Box$

The closest similar point problem is difficult in that there can be images with
$n/2$ labels where each occurs exactly twice, in which case the problem is
essentially the same as sorting, requiring $\Omega(n^{3/2})$ total energy in the worst case.
However, when the only labels are black or white the problem becomes
considerably easier.
The tradeoff of number of colors vs. total energy isn't examined here.

\begin{figure}[t]

\centerline{\includegraphics{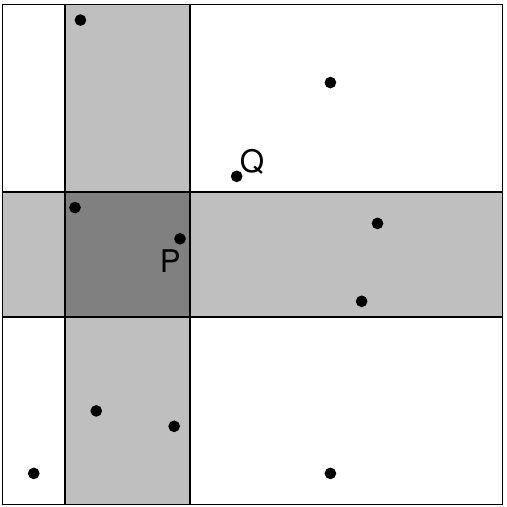}}

\caption{The closest point to P is Q}

\hrulefill

\label{fig:l2nearestpoint}
\end{figure}

\begin{theorem} \label{thm:closestsimilar12infinity}
Using peak power $s$, $1 \leq s \leq n$, in $\Theta(n^2/s)$ time the closest
similar point problem can be solved for a black/white labeled image, where the
metric is $\ell_1$, $\ell_2$, or $\ell_\infty$.
\end{theorem}
\textit{Proof:\,}
For the $\ell_1$ and $\ell_\infty$ metrics the problem was solved in
Theorem~\ref{thm:closestblack1infinity}.  For the $l_2$ norm a somewhat more
complex algorithm is used, closely following that in \cite{MillerStout89}.
Figure~\ref{fig:l2nearestpoint} helps illustrate the approach.  Suppose $p$
is a black pixel within the dark subsquare, and suppose the closest black point
has been found within the union of the subsquare and the horizontal and vertical bands
(the lighter gray regions). The only way there might be a
closer black pixel is if $p$ is closer to a corner $c$ than it is to any black
pixel found so far, for in this case there might be a black pixel $q$ in the
white quadrant corresponding to $c$ that is the closest one to $p$.  Call $p$ 
a \textit{special point} if it satisfies this criterion.
An important fact is that there are at most 2 points within the square which are
closer to
$c$ than to any point found so far.  If there were 3 or more such points, one
would be closer to another than to the corner (see~\cite{MillerStout89}).  Thus
in total there are at most 8 special points for which white regions need to be
considered.

To start the process, each squirrel is assigned to a subsquare of size $n/s$.  These
are the black subsquares in Figure~\ref{fig:l2nearestpoint}.  In linear time,
for each black
point it locates the nearest black, if any, in the square.
(It can do this by, say, simulating the recursive
approach described for the entire image.)\,
At this stage the only points which are not guaranteed to have found the
closest are those points which are the leftmost or rightmost within a row, or
highest or lowest in a column, of the subsquare.  The only points within the
banded regions which might be closest to them are the leftmost ones in each row
in the banded region to the right of the square, the rightmost ones in each row
in the banded region to the left of the square, and similarly for the vertical
banded region.   Row- and column-wise sweeps as in
Theorem~\ref{thm:closestblack1infinity} can be used to simultaneously find the
appropriate banded region points for all subsquares, and finding the points
within the subsquare is similar.   There are at most $\Theta(n/\sqrt{s}\,)$
points within the square that have to consider at most $\Theta(n/\sqrt{s}\,)$
points within the banded region, so simple comparisons of all of the inside
points with all of the outside ones can be done in $\Theta(n^2/s)$ time.  Then
the special points are located.  A traversal along the row
corresponding to the top row of the square is performed,  where at each column the
distance from the special point to the lowest black pixel above that row is
computed, and if this is closer than any point found so far then it is kept.
Similar operations are performed along the horizontal and vertical bounding
lines in all directions.  The traversal takes
$\Theta(n)$ time, and when all of the traversals are completed the
closest black neighbor of each black pixel has been found.  Then the same
algorithm is applied to the white pixels, locating the closest white.
$\Box$

Note that the above approach does not directly solve the nearest black point
problem for the $\ell_2$ metric because there may be more than 8 special points.
For example, there may be no black pixels in the square or banded regions, and
hence every pixel in the square is special.

\section{Minimizing Total Energy Used by Any Single Processor}
  \label{sec:singleprocessor}

In some applications, such as sensor networks, an important consideration is
the maximum energy used by any sensor.  This is because the sensors are assumed
to have their own, limited, power, as opposed to the externally supplied power
which motivated this work.

Recall that the power used by a processor corresponds to the number of times it
was visited by a squirrel.  For the preceding algorithms,
in most steps no processor is visited more then
$\Theta\left(\frac{1}{n^2} \cdot \mathrm{(Total~Power)}\right)$
times, i.e., no processor had power requirements more than the average.
However, when data was collected and moved to a subsquare for processing by a
standard mesh-connected computer algorithm then this was not true, since
processors in the subsquare were on continuously during this step, far more
than their share. For example, in the last stage of recursion in the image
component algorithm for Theorem~\ref{thm:imagelabel}, 
the processors simulate a standard mesh algorithm taking $\Theta(n^{1/4})$ time,  and
hence expend $\Theta(n^{1/2})$ energy, but the average per processor 
for the entire algorithm is only $\Theta(\log n)$.

\begin{figure}[t]

\centerline{\includegraphics{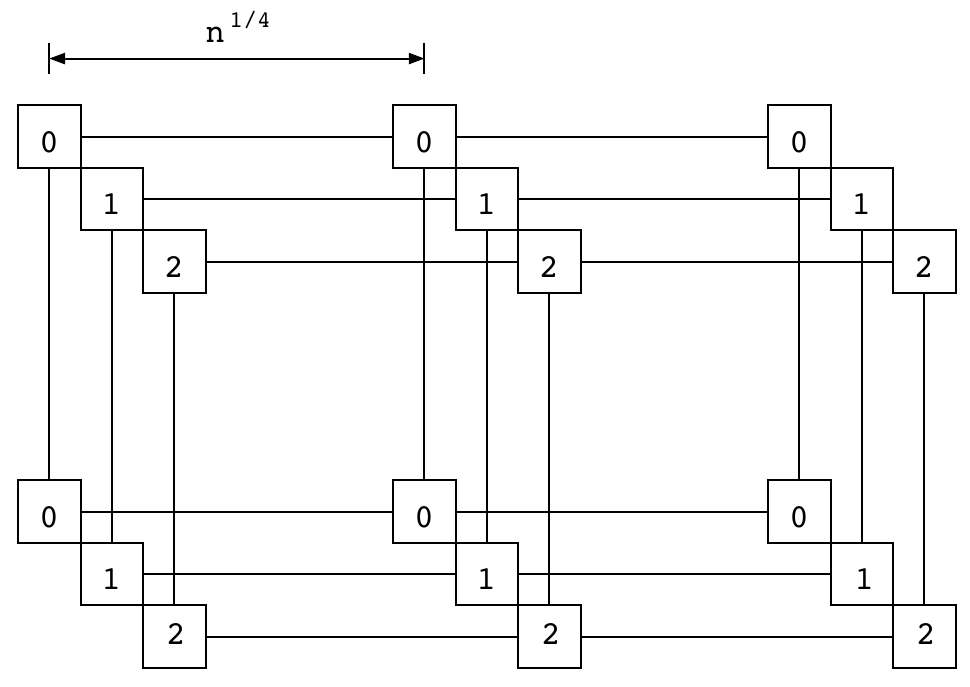}}
\medskip

\centerline{Numbers indicate when the processor is used to simulate the
standard mesh algorithm}

\caption{Expanding the simulating mesh}

\hrulefill

\label{fig:expandmesh}
\end{figure}

However, these power-intensive steps can be modified so that the average power
per processor is still a constant.  To do so, the $n^{1/2} \times n^{1/2}$
submesh is expanded, as in Figure~\ref{fig:expandmesh}, so that the simulating
processors are $n^{1/2}$ apart.  Then a fixed number of steps of the simulation
are performed, where each step now takes $\Theta(n^{1/2})$ time. After this,
the location of the simulating processors are moved diagonally 1 step, as
indicated in the figure, and then another fixed number of steps are simulated,
and so forth.  The number of steps is chosen so that the simulated algorithm is
finished by the time the diagonal movement would place simulating processors on
top of ones previously used.

It is easy to see that now no processor is used more than a constant number of
times during the simulation, either for calculation or as part of a communication
path.  While the notion of grouping processors together
and having only a few be active at one time has been used in sensor networks
(e.g.,~\cite{Span01,SiBaPrasanna04}), it is unusual to do this when it significantly
increases the time.  The time has increased by a factor of
$n^{1/4}$, but this merely makes it equal to the time needed to move data to
the subsquare.  Thus it does not increase the total time by more than a
constant multiple.    Similar changes can be made concerning  the
use of the leftmost column in
Theorem~\ref{thm:imagereduce}.

Summarizing, we have the following:

\begin{theorem}
Using the indicated changes, all of the preceding algorithms can be modified so
that the peak power, total power, and time do not change by more than a
constant multiple, and simultaneously each processor uses only
$\mathrm{O}\left(\frac{1}{n^2} \cdot \mathrm{(Total~Energy)}\right)$ energy.
$\Box$
\end{theorem}

\section{More Power, Faster Computation} \label{sec:morepower}

The previous sections considered the time/power tradeoffs when the power was $O(n)$, showing that the time achieved can be within a log factor of the time used by a standard mesh algorithm where all processors are active all the time.
Those algorithms kept the power per active processor a constant.
However, a different approach, useful when the power $p(n)$ is $\omega(n)$, is to use it to increase the speed of the active processors.
We use a model where a processor with power $p \geq 1$ can run at speed $\Theta(\sqrt{p})$, though the approaches can extended to models where the speed is $\Theta(p^c)$ for some $0 < c$.

The simplest use of the algorithms is to allocate $p(n)/n$ power to the algorithms when $n$ squirrels are being used.
Let $T(n)$ be the time of the original algorithm when $n$ squirrels are in use.
Then the new version, with faster squirrels, will take time $\Theta\left(T(n) \sqrt{n/p(n)}\right)$.
In particular, if the power is that of the standard mesh model, i.e., $\Theta(n^2)$, we have

\begin{proposition} \label{prop:labeledfigures}
If the power is $\Theta(n^2)$ then for an $n \times n$ image, in $\Theta(\sqrt{n})$ time
\begin{enumerate}
\item If the image has strongly labeled figures, then broadcast and reduction over figures
can be performed, and for each figure one can determine
if it is contained in any others, obtain the label of the smallest container,
and determine its nesting level.

\item If the image is a black/white image then for each figure one can
determine the smallest container of the same color, and the  the closest
similar point problem can be solved, where the
metric is $\ell_1$, $\ell_2$, or $\ell_\infty$.

\item If each pixel has a label then the closest
black point problem, and the closest differing point problem, can be solved for
the $\ell_1$ and $\ell_\infty$ metrics.

\end{enumerate}
$\Box$

\end{proposition}

A slightly more sophisticated use of the power can improve the time of the component labeling approach
in Theorem~\ref{thm:imagelabel}.
The above would give $\Theta(\sqrt{n} \log n)$, but the log factor can be eliminated.
At each stage the component label information of the boundaries of 4 $k \times k$ images are combined and the result is the component label information of the boundary of a $2k \times 2k$ image, for some $k$ a power of 2.
This results in 2 times as many words of information as in each of the smaller image, where each word needs to be moved twice as far as the words of the original 4 images.
The power per word is proportional to the area of the square it comes from, so the power for the words from the larger square is twice as much as the power per word from the original squares, and hence each word from the larger square can be moved $\sqrt{2}$ times as fast, taking $\sqrt{2}$ times as long to reach its destination.
Conceptually this is like having many small squirrels at the initial stages, and they combine to become fewer, faster squirrels as the stages progress.
The time per stage increases geometrically and hence the total time is proportional to the time of the last stage, which is $\Theta(\sqrt{n})$.
This gives

\begin{proposition} \label{prop:labelthefigures}
If the power is $\Theta(n^2)$ then for an $n \times n$ image the figures can be labeled in $\Theta(\sqrt{n})$ time.\\
$\Box$
\end{proposition}

\noindent An immediate corollary is that under the same power conditions the maze problem (see Corollary~\ref{cor:maze})  can be solved in $\Theta(\sqrt{n})$ time.

For all of the problems mentioned in Propositions~\ref{prop:labeledfigures} and~\ref{prop:labelthefigures} the standard mesh algorithms also use power $\Theta(n^2)$ but require time $\Theta(n)$.

\section{Final Remarks} \label{sec:final}

This paper has been concerned with minimizing peak power usage by
mesh-connected computers, or with combining the resulting algorithms with higher power to achieve
algorithms which are faster than the classical mesh algorithms yet use the same power.
Our time analysis of the use of higher power assumed that the speed increased as the square root of the power, but similar (though potentially far weaker) results are possible whatever the increase is.
One subtlety is that some may prefer a model where message transmittion speed increases as the square root, but that the time for each calculation stays the same.
However, the times of the algorithms given in Section~\ref{sec:morepower} would remain the same.

The algorithms
herein were only given for 2-dimensional meshes, especially since images are
naturally 2-dimensional, but one can use similar approaches to
develop power reducing algorithms for higher dimensional meshes.  However, as
the dimension increases there is a smaller range in which to lower peak power
without increasing the time required.  For example, in a 3-dimensional mesh of
$n$ processors, summing values from each processor can be done in
$\Theta(n^{1/3})$ time, and any algorithm which achieves this minimal
time must have a peak power of $\Omega(n^{2/3})$.  In contrast, in 2 dimensions
the minimal time is $\Theta(n^{1/2})$, which can be achieved with a peak power
of $\Theta(n^{1/2})$.

A more optimistic viewpoint is that for the same peak power
there are many problems that a
3-d mesh can solve faster than a 2-d one, just as 2-d meshes can be superior
to 1-d meshes. Sorting is one such example. Unfortunately, for most of the problems in this
paper that is not possible since the algorithms provide work-optimal,
or nearly work-optimal,
tradeoffs compared to serial algorithms. 
However, this is only true for the range of peak power considered in the theorems.
If a larger peak power is used then the 2-dimensional algorithms are no longer work-optimal
and there is the possibility of algorithms for 3-dimensional meshes being work-optimal.

For 2-dimensional meshes it seems that basic matrix operations are not amenable
to any peak power reduction without increasing the time.
For example, it appears that multiplying $n \times n$ matrices requires $\Omega(n^3)$ total energy on a 2-dimensional mesh, despite algorithms such as Strassen's which reduce the serial work to $o(n^3)$. 
However, one can show that on a 3-dimensional mesh it is possible to 
multiply matrices using $\Theta(n^2)$ peak power and $\Theta(n^{2/3})$ time.
For some problems it isn't clear what effect the dimension has.
For example, to find the median, the peak power vs.\ time tradeoffs are unknown
for arbitrary dimensions.
Presumably the optimal tradeoff for dimension 2 would point to the tradeoff for arbitrary dimensions.

For graphs, in general the more structured the input the easier it is to reduce the power.
For example, adjacency graphs yield faster algorithms than do unordered edges.
Part of the explanation for this is that less global rearrangement of the data
is needed, an operation which is quite power-intensive.  In some cases one
might need to do an initial power-intensive operation with restricted peak
power, in which case the time will increase.  However, there might be efficient
algorithms for properly organized data.  For example, some
geometry problems on point data can be solved significantly faster if the points
have been sorted by x-coordinate.   Thus if one is solving a sequence of such
problems it may be useful to organize the data initially and view the 
organizational time as being amortized over subsequent operations.

It is interesting to note that some of algorithms, such as in Theorem~\ref{thm:closestblack1infinity}
 use a fixed pattern of
power usage, in that the time at which a processor is active is independent of the
data. The messages passed, of course, do depend on the data.  One 
exception is the depth-first search used by individual squirrels to label
figures in a subsquare at the start of the algorithm in Theorem~\ref{thm:imagelabel}. 
They could have
used the fixed activation pattern employed by the subsequent stages of the
labeling process, but an extra logarithmic factor would have been introduced.
It is unclear what the optimal time is for a single squirrel if its pattern of
motion must be independent of the data.

\end{document}